\documentclass[a4paper,11pt]{article}
\pdfoutput=1
\usepackage{graphicx}
\usepackage{bm}
\usepackage{epsf}
\usepackage{rotating}
\usepackage{float}
\usepackage{epsfig,graphics,rotate,color}
\usepackage{color}
\usepackage{array}
\usepackage[utf8]{inputenc}
\usepackage[T1]{fontenc}
\usepackage[caption=false]{subfig}
\usepackage[bottom]{footmisc}
\usepackage{adjustbox}
\usepackage{multirow}
\usepackage{subcaption}
\usepackage[utf8]{inputenc}
\usepackage{amsmath}
\usepackage{amsfonts}
\usepackage{amssymb}
\usepackage{lineno}

\pdfoutput=1 
\usepackage{jinstpub} 

\newcommand{\beq}{\begin{eqnarray}}
\newcommand{\eeq}{\end{eqnarray}}

\def\cdwater{200}  
\def\cdvinegar{50} 
\def\cdsugar{80} 
\def\cdsoda{20} 

\def\PL{450} 
 
\def\excite{340} 

\def\volume{400} 
\def\qe{25} 
\def\qee{3} 
\def\voltage{1150} 
\def\gain{1.1\times 10^7} 
\def\gaine{10} 

\def\nphoto{70}
\def\nphotoe{20}
\def\zerofrac{10} 
\def\trigger{100} 
\def\photol{15}
\def\photom{35}
\def\photoncut{3} 
\def\datapC{10} 
\def\resolpmt{15}
\def\resolflour{15}
\def\resolscat{15}

\title{Cooking Carbon Dots – Making an Instant Neutrino Detector in Your Kitchen}
\author{
D.~W.~King, 
K.~Samokovlisky, 
D.~Panova, 
A.~Dimitrichenko, 
L.~Umrikhin, 
T.~Katori, 
and A.~Rakovich 
}
\affiliation{Department of Physics, King's College London, WC2R 2LS London, UK.}

\emailAdd{teppei.katori@kcl.ac.uk, aliaksandra.rakovic@kcl.ac.uk}

\abstract{Liquid scintillators underpin a wide range of radiation detectors, including those used in neutrino physics, but typically rely on organic fluors dissolved in hazardous and costly solvents. Here, we show that carbon dots -- nanoscale fluorescent carbon materials -- synthesised from simple household ingredients using a microwave can function as water-based liquid scintillators~\cite{Zhao:2024azj}. These carbon dots dispersed in water  produce light yields up to $\nphoto\pm\nphotoe$ photons per MeV and enable the detection of atmospheric muons. This yield is sufficient to detect low-energy protons in water Cherenkov neutrino detectors, expanding their programs in both particle physics and astrophysics. These results establish an accessible, low-cost and environmentally benign route to scintillator development, opening new opportunities for large-scale  radiation detection.
}


\begin{document}
\maketitle
\flushbottom

\section{Introduction
\label{sec:Introduction}
}

Liquid scintillator detectors play a central role in neutrino physics. Large-scale liquid scintillator detectors, such as KamLAND~\cite{KamLAND-Zen:2024eml}, SNO+~\cite{SNO:2025chx}, and JUNO~\cite{JUNO:2025gmd} operate over a broad energy range, detecting neutrinos from low-energy solar, reactor, and geo sources to high-energy atmospheric neutrinos. These detectors utilize large volumes of liquid scintillator monitored by an array of large photo-cathode photo-sensors. In contrast, the NOvA experiment~\cite{NOvA:2025tmb} adopts a segmented design, in which liquid scintillator is contained within PVC (polyvinyl chloride) cells and wavelength shifting fibers read the light outputs from a high-energy accelerator and atmospheric neutrino interactions. 

In all of the above cases, the scintillation medium is based on organic fluors dissolved in organic solvents. Such systems dominate radiation detection owing to their favourable properties. Firstly, the light output achievable with organic fluors is high and is suitable for detecting low-energy events such as neutrino interactions. Secondly, the emission is in the $\sim$ns timescale, and emission in the blue spectral region matches typical photon sensor sensitivities. Thirdly, they are stable for long exposure measurements. However, the use of organic solvents introduces significant drawbacks, including flammability and environmental concerns, which constrain their broader application.

Water-based liquid scintillators offer a promising alternative for next-generation neutrino detectors such as THEIA~\cite{Theia:2019non}. The feasibility of this approach in neutrino physics has been demonstrated by the ANNIE experiment at Fermilab~\cite{ANNIE:2023yny}, with further dedicated R\&D efforts being currently in operation, including EOS at LBNL ~\cite{Anderson:2022lbb} and BNL 1~ton detector~\cite{Xiang:2024jfp}, and others expected to join this activity in the near future, e.g. the Button detector at Boulby Lab~\cite{Bhattacharya:2025qwl}. These systems typically rely on surfactants to enable dispersion of organic scintillating components in water~\cite{Yeh:2011zz}.

Carbon dots are a class of nanoscale carbon-based fluorescent materials, typically comprising of small carbon cores with chemically active surfaces. Since their initial discovery, carbon dots have attracted considerable attention due to their strong and tunable photoluminescence, chemical stability, and straightforward synthesis from a wide range of precursors, including simple organic compounds~\cite{AngewandteBaker2010,C4CS00269E}. Their optical response is generally attributed to a combination of surface-related states, structural defects, and synthesis-dependent fluorescent species, leading to broad absorption features and emission that depends sensitively on preparation conditions~\cite{C8TB00153G,D1NA00559F,C5CC07754K,JPCLKumbhakar2017}. Carbon dots form stable colloidal dispersions in polar solvents, including water~\cite{D2RA07180K}, without the need for additional surfactants, which has motivated their application in areas such as bioimaging, sensing, photocatalysis, and optoelectronics~\cite{C3RA47994C,C4CS00269E}.

\begin{figure}[t!]
 \begin{center}
\includegraphics[width=0.5\columnwidth]{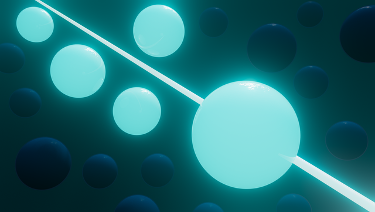}
 \end{center}
\vspace{-2mm}
\caption{Artist's impression of carbon dots excited by an atmospheric muon emitting blue light.}
\label{fig:cartoon}
\end{figure}

Water-dispersible nanoparticles, such as hydrophilic semiconductor quantum dots~\cite{Zhao:2024azj}, have recently been proposed as components of water-based liquid scintillators. However, many such materials are based on semiconductor compositions containing heavy metals, which can pose environmental and health concerns; in addition, their synthesis can present challenges in cost and large-scale deployment. Carbon dots provide a complementary alternative: they can be synthesised from inexpensive precursors, are environmentally benign, and are readily scalable to the large target masses required for neutrino detectors, typically at the kiloton scale. Their compatibility with water as a solvent further motivates their use in water-based liquid scintillators (Fig.~\ref{fig:cartoon}), offering the possibility of combining scintillation with established water Cherenkov detector technologies. The potential of carbon dots for radiation detection~\cite{photonics12090854,doi:10.1021/acs.nanolett.5c05502}, including applications in neutrino detectors, is investigated here.

\section{Carbon dots
\label{sec:carbon}
}

\subsection{Synthesis of carbon dots}

\begin{figure}[t!]
 \begin{center}
\includegraphics[width=0.42\columnwidth]{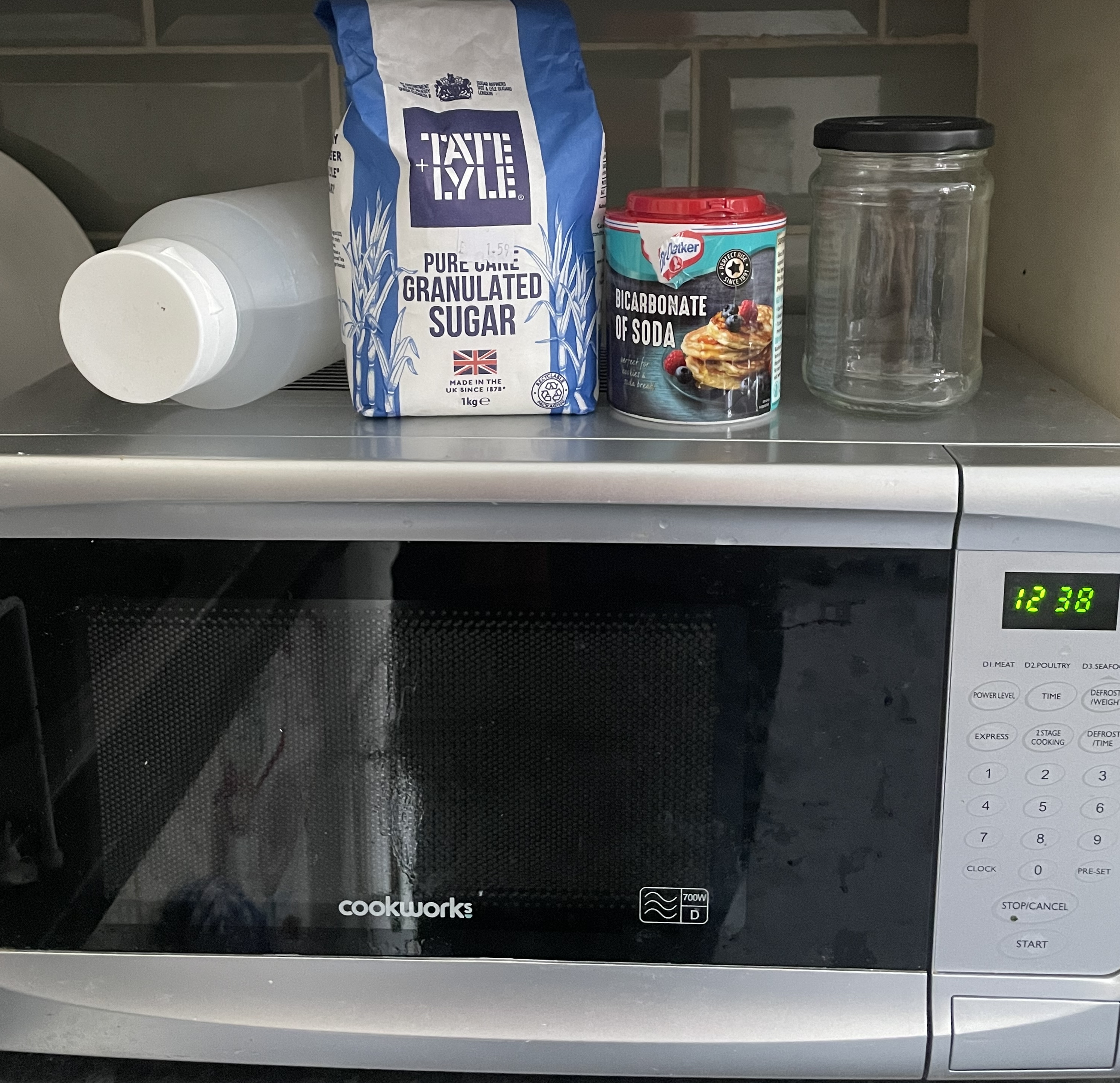}
\includegraphics[width=0.54\columnwidth]{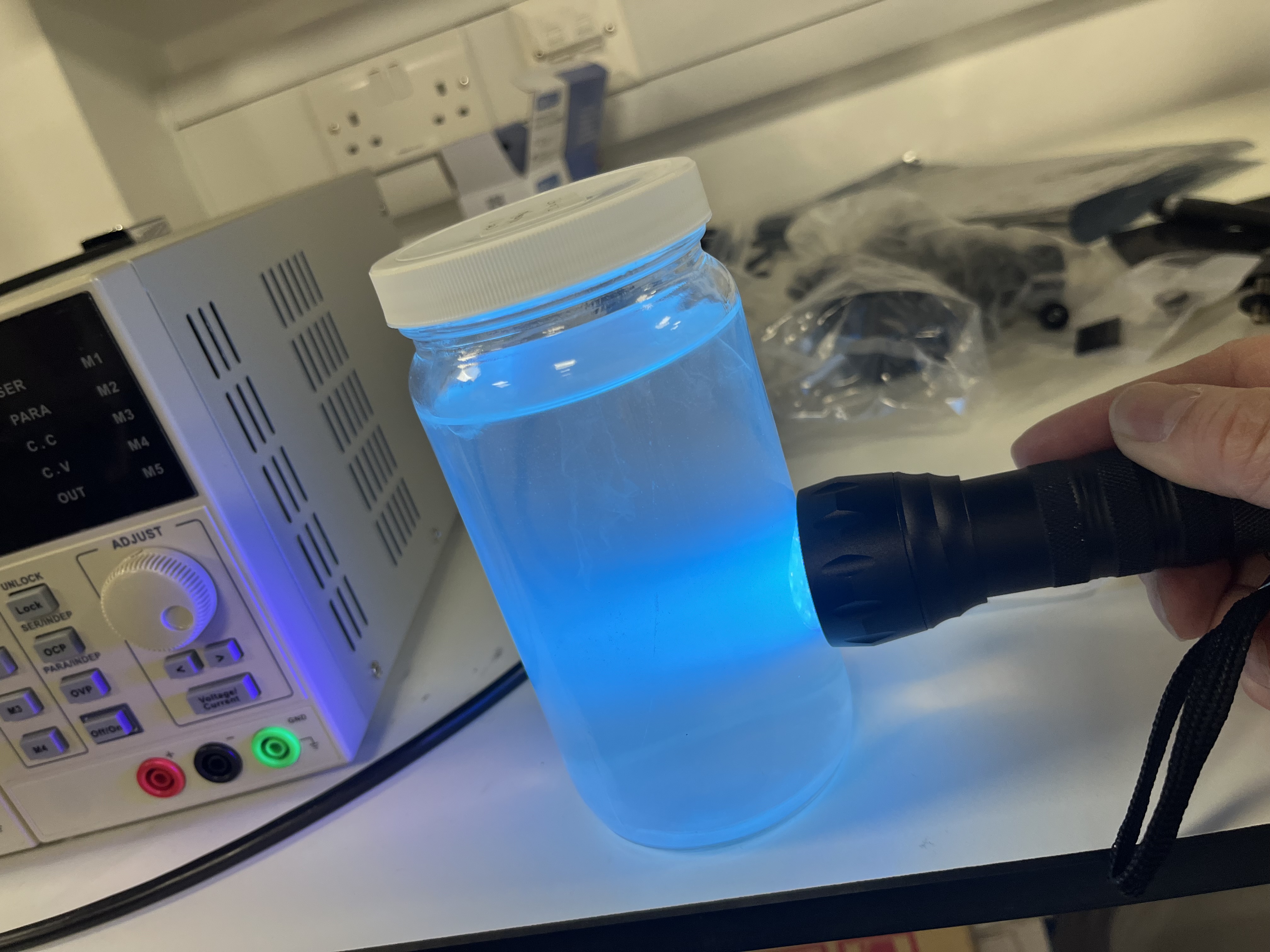}
 \end{center}
\vspace{-2mm}
\caption{Left, ingredients of carbon dots: sugar, vinegar, and baking soda. Right, carbon dots are diluted in water until the blue emission becomes visibly the brightest.}
\label{fig:production}
\end{figure}

Carbon dots were synthesized using a microwave-assisted synthesis~\cite{StrangelyAmusing}, selected for its simplicity and accessibility. The precursors and the microwave used are shown in Fig.~\ref{fig:production} (left). The reaction solution was prepared by combining tap water ($\cdwater$~mL), distilled white vinegar ($\cdvinegar$~mL), and white granulated sugar ($\cdsugar$~g) in a glass jar. The solution was stirred to dissolve the precursors and then heated in a microwave oven for 5~minutes at intensity of 800~W. Under these conditions, sucrose undergoes inversion to glucose and fructose, while the acidic environment provided by the vinegar accelerates the reaction. Upon completion of the microwave step, reaction solution was cooled to room temperature, and baking soda ($\cdsoda$~g) was gradually added under stirring to neutralise it.

The result of the synthesis was a dark liquid, which was diluted with tap water at approximately 1 to 70 ratio to produce $\sim\volume$~mL of carbon dot dispersion for scintillation measurements. The optimum concentration for the measurements was obtained by achieving the strongest blue fluorescence emission by eyes from the carbon dots dispersion under UV illumination by a UV torch, Fig.~\ref{fig:production} (right). Carbon dots samples were independently prepared by several students in a home setting; two representative samples were selected for further study. Commercially-available carbon dots, purchased from Sigma-Aldrich~\cite{Aldrich} were used to benchmark the responses of the home-made samples.

\subsection{Carbon dots properties}

\begin{figure}[t!]
 \begin{center}
\includegraphics[width=0.44\columnwidth]{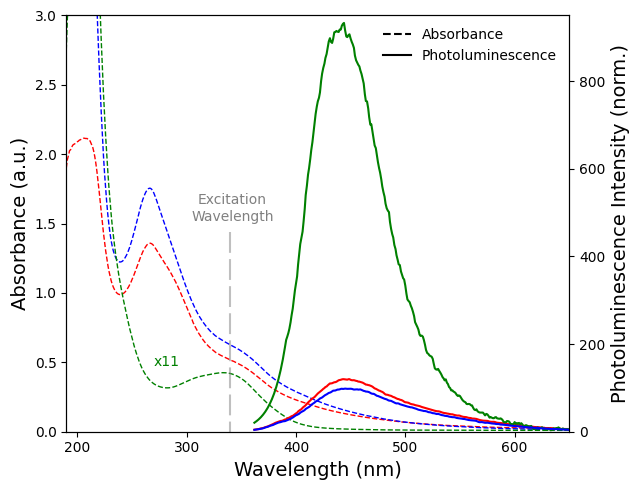}
\hspace{2mm}
\includegraphics[width=0.41\columnwidth]{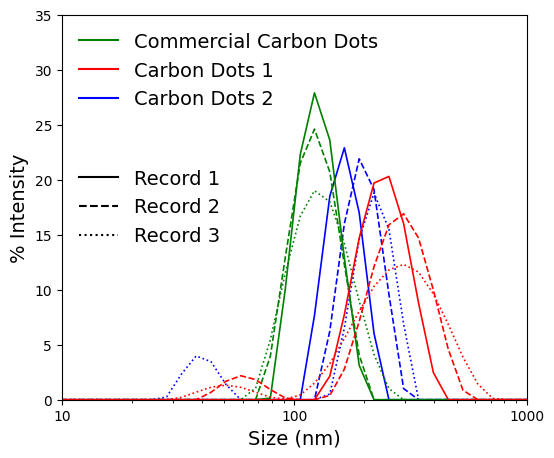}
 \end{center}
\vspace{-2mm}
\caption{Left, absorbance measurements (dashed) used to normalize photoluminescence (solid) data (excitation wavelength = $\excite$~nm, vertical dashed line). Right, the hydrodynamic size of the three solutions measured with the dynamic light scattering method, in triplicate (solid, dashed, and dotted lines). Carbon dots sample 1 in red, Carbon dots sample 2 in blue, and a commercial carbon dots sample, in green.} 
\label{fig:optics}
\end{figure}

The optical properties of the carbon dots dispersions were characterised using absorbance and photoluminescence spectroscopies~\cite{Shimadzu,CaryEclipse}. An excitation wavelength $\excite$~nm was selected from the absorbance spectra for photoluminescence measurements, such that it lays within an absorption peak of each sample, and was separated by 20~nm from the onset of the photoluminescence detection window. Photoluminescence spectra were recorded and then normalised to the absorbance of each sample at the excitation wavelength. The resulting emission spectra are shown in Fig.~\ref{fig:optics}, left. As can be seen from this figure, all three samples exhibited emission peaks at approximately $\PL$~nm. The home-made samples (red and blue) had substantially lower emission intensity compared to the commercially-procured sample (green), after absorbance normalisation, which is consistent with a lower photoluminescence quantum yield for these samples. On the other hand, all samples exhibited emission spectra with similar shape and peak position, suggesting a common emission origin, likely associated with similar emissive states. The observed differences in intensity reflects variations in their relative population or radiative efficiency, potentially arising from differences in preparation and processing between the samples.

The hydrodynamic size distributions of the carbon dot solutions were measured using dynamic light scattering~\cite{Malvern}  (Fig.~\ref{fig:optics}, right). 
The home-made samples (red and blue) exhibited multimodal size distributions across repeated measurements, with a peak centred around $\sim$40-60~nm and a second, broader peak in the 200–300~nm range. The smaller peak is consistent with individual nanoparticles, while the larger size population suggests the presence of aggregates or other larger structures in solution. Furthermore, the relative prominence of these peaks varied between measurements, indicating significant sample variability. In contrast, the commercially available sample (green) showed a consistent, single-peak size distribution centred around $\sim$120~nm, indicating a more uniform colloidal population than the home-made samples.

In addition, the home-made samples showed visible filament-like residue over several days, suggesting limited long-term stability of the dispersions. This behaviour may arise from several processes associated with bottom-up carbon dot synthesis from molecular precursors. Incomplete carbonisation of sugar-derived precursors is known to yield residual polymeric or carbonaceous species, which can evolve into extended network-like structures over time~\cite{TITIRICI2015325,He2021}. Furthermore, the colloidal stability of carbon dots depends sensitively on surface chemistry, pH, and ionic conditions, and the absence of post-synthesis purification can promote aggregation and phase separation in solution~\cite{C4CS00269E,C5CC07754K,D2RA07180K}. Residual organic species formed during synthesis may also undergo slow crystallisation or further polymerisation, contributing to the formation of filamentous or sedimented structures~\cite{Aydincak2012,Sevilla2009}. In the present case, no filtration or purification steps were applied, and all of these processes may contribute to the observed temporal evolution of the samples.

\section{Carbon dots water-based liquid scintillator radiation tests
\label{sec:cosmic}
}

\subsection{Setup}
\begin{figure}[t!]
  \begin{center}
    
\includegraphics[width=0.60\columnwidth]{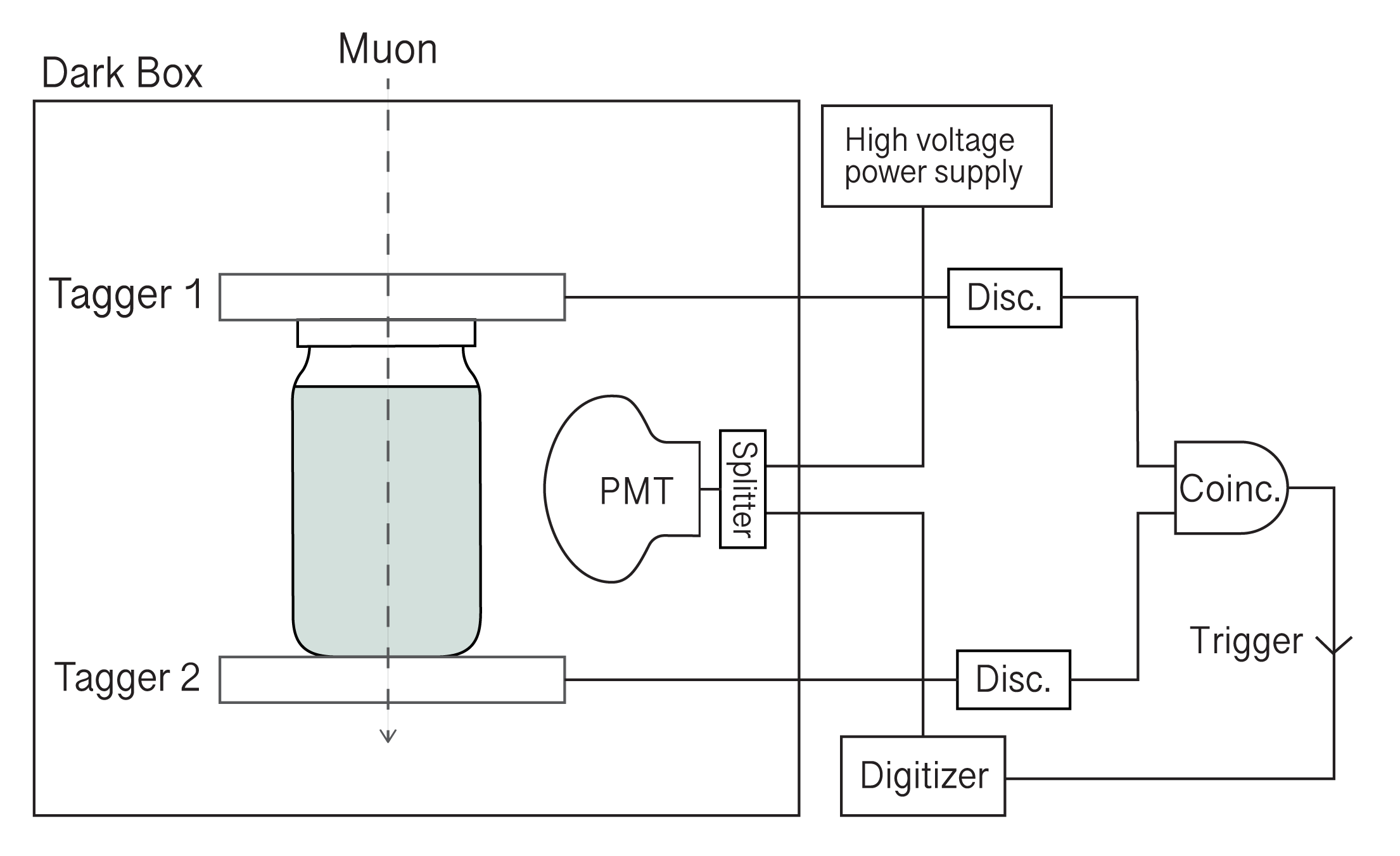}
\hspace{2mm}
\includegraphics[width=0.32\columnwidth]{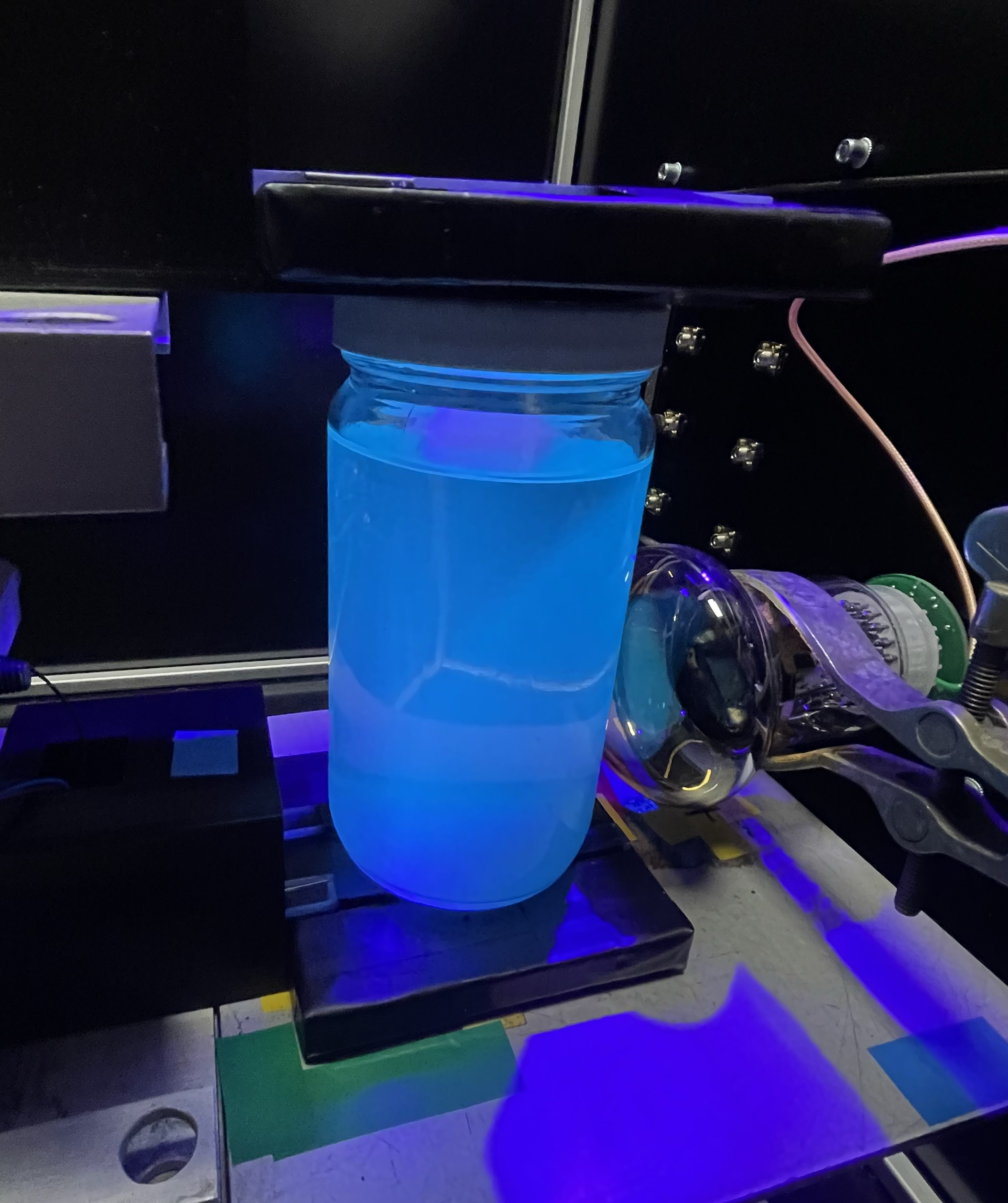}
 \end{center}
\vspace{-2mm}
\caption{Left, schematics of the atmospheric muon response measurement. Muons were tagged by two plastic scintillators. After discrimination, a coincidence signal was produced to trigger the digitizer, and PMT responses were measured. Right, a photograph of the inside of the dark box. The carbon dots solution is illuminated by a UV torch.}
\label{fig:setup}
\end{figure}

 We used a similar setup to one employed in a previous study~\cite{Zhao:2024azj}, shown in Fig.~\ref{fig:setup} (left).
 Inside a dark box, a glass jar containing a $\sim\volume$~mL sample was placed between two cosmic ray taggers, above and below. 
 Each tagger was made from a 10~cm$~\times~$10~cm$~\times$~1~cm plastic scintillator~\cite{Eljen} and a silicon photo-multiplier~\cite {SiPM}. The signals from cosmic ray taggers were discriminated, and a coincidence unit was used to produce the trigger. 
 A positive high-voltage 3-inch photomultiplier tube (PMT) ~\cite{NNVT} was mounted on a mechanical stand, facing the photocathode of the PMT close to the jar's side, to measure emission from carbon dot samples, and the digitised PMT signals were recorded~\cite{CAEN}. 

\subsection{Data}
\label{sec:data}

\begin{figure}[t!]
 \begin{center}
 \includegraphics[width=0.54\columnwidth]{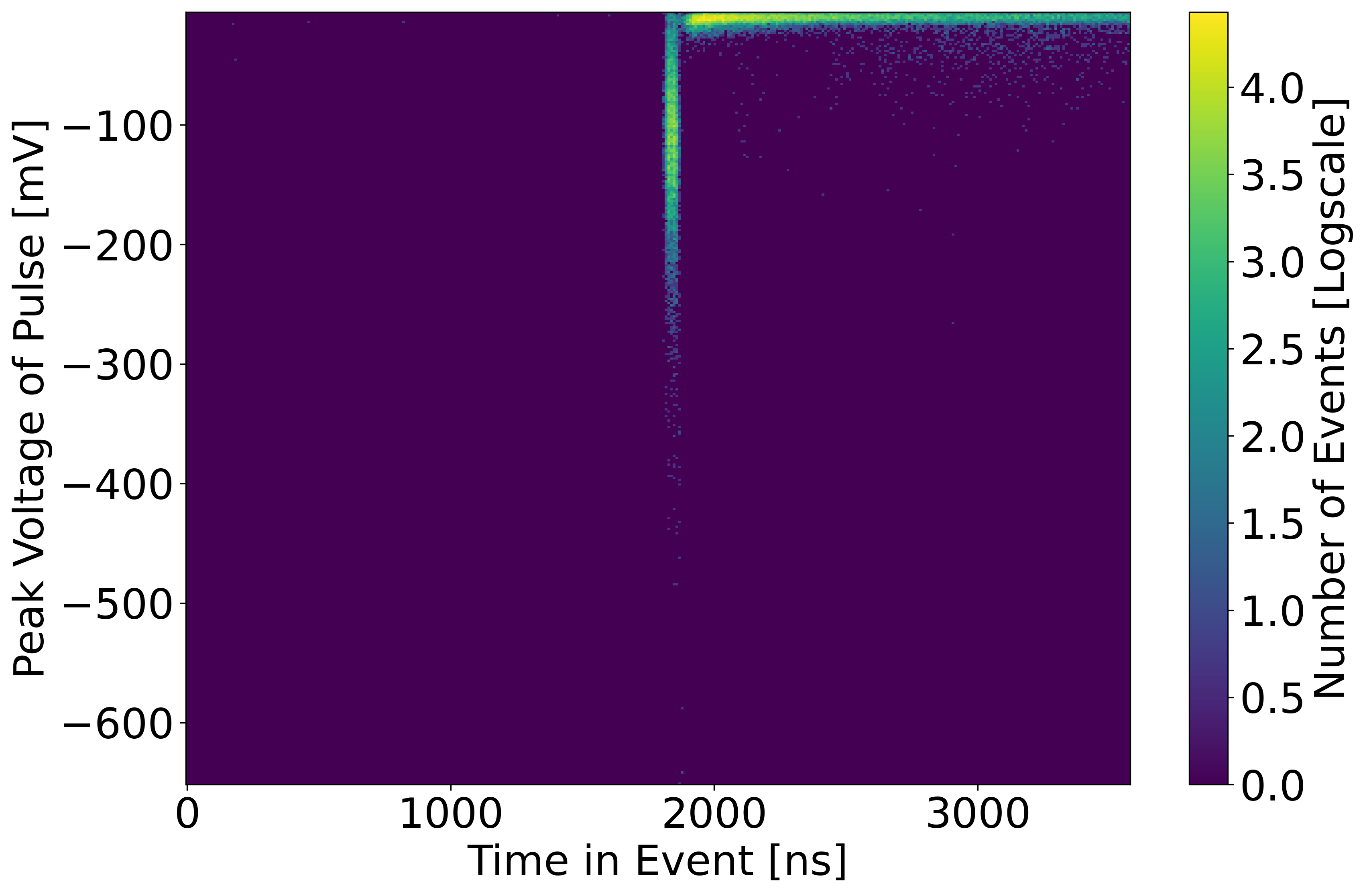}
\includegraphics[width=0.45\columnwidth]{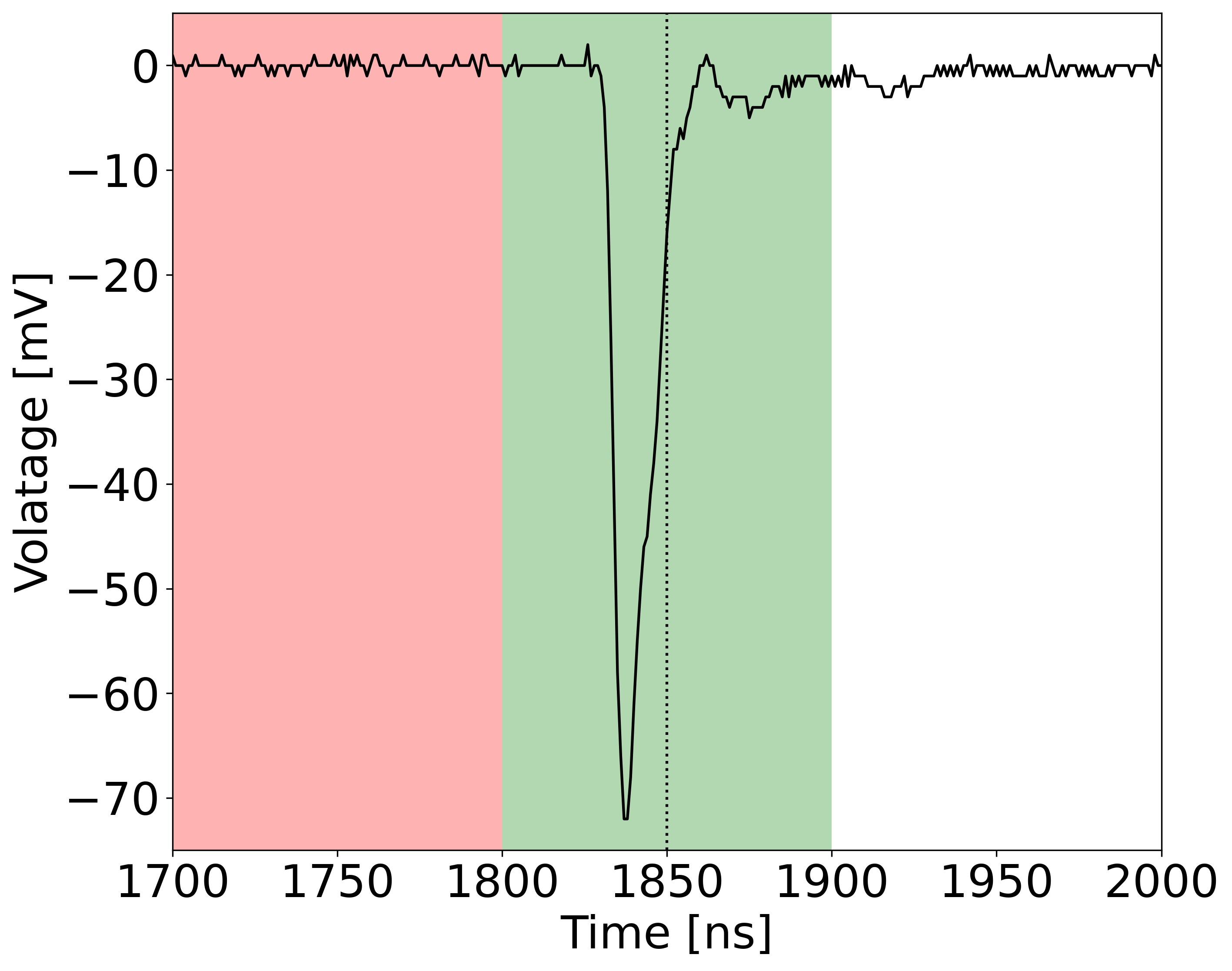}
 \end{center}
\vspace{-2mm}
\caption{Left figure shows the two-dimensional histogram of the pulse peak height (mV) vs. peak pulse time (ns). Count is shown with color in logarithmic scale. Right, a typical pulse observed by a 3-inch PMT from the carbon dots muon response. Dashed line indicates the expected arriving time of the pulse from the known delays of cables and electronics. The pink region is used to calculate the baseline. The 100~ns green region is used to integrate the pulse to calculate the charge. }
\label{fig:data}
\end{figure}

Fig.~\ref{fig:data} shows the recorded data. Since the taggers are slightly bigger than the jar cross section, a small fraction of tagged muons did not hit the carbon dot sample, as a result of which about $\sim\zerofrac$\% of tagged events did not contain PMT signals.  
Additionally, variations in the trajectories of muons passing through the jar led to differences in their track length and, consequently, in the energy deposited per event, producing a broad charge distribution.
We operated this PMT with a gain of $\gain$ at $+\voltage$~V. A gain calibration with an LED~\cite{LED} was performed with Poisson distribution for photon statistics. From multiple measurements and different analyses, an error of $\gaine$\% was assigned to the gain of this PMT.

Fig.~\ref{fig:data}, left, shows the two-dimensional histogram of pulse height \textit{vs.} pulse time. Although most of events were at the expected arriving time, there were secondary after pulses which were not included in this analysis. To analyze the data, the pulse was integrated to find the charge of each event whilst avoiding noise. First, a $\trigger$~ns signal window was set from the known delays of signals (area shaded in green in Fig.~\ref{fig:data}, right). 
Second, the data recorded before the signal window was used to calculate the baseline (pink shaded area, Fig.~\ref{fig:data}, right). Third, we calculated the charge by integrating $\trigger$~ns window for each pulse above the baseline. To extract further information, these results were compared to a simulation of the experiment.

\subsection{Simulation}
\begin{figure}[t!]
 \begin{center}
\includegraphics[width=0.35\columnwidth]{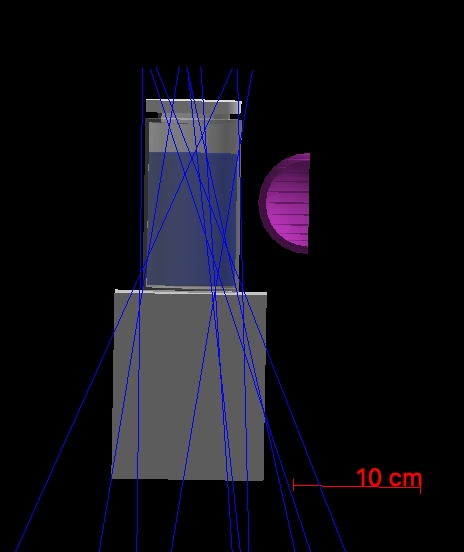}
\includegraphics[width=0.64\columnwidth]{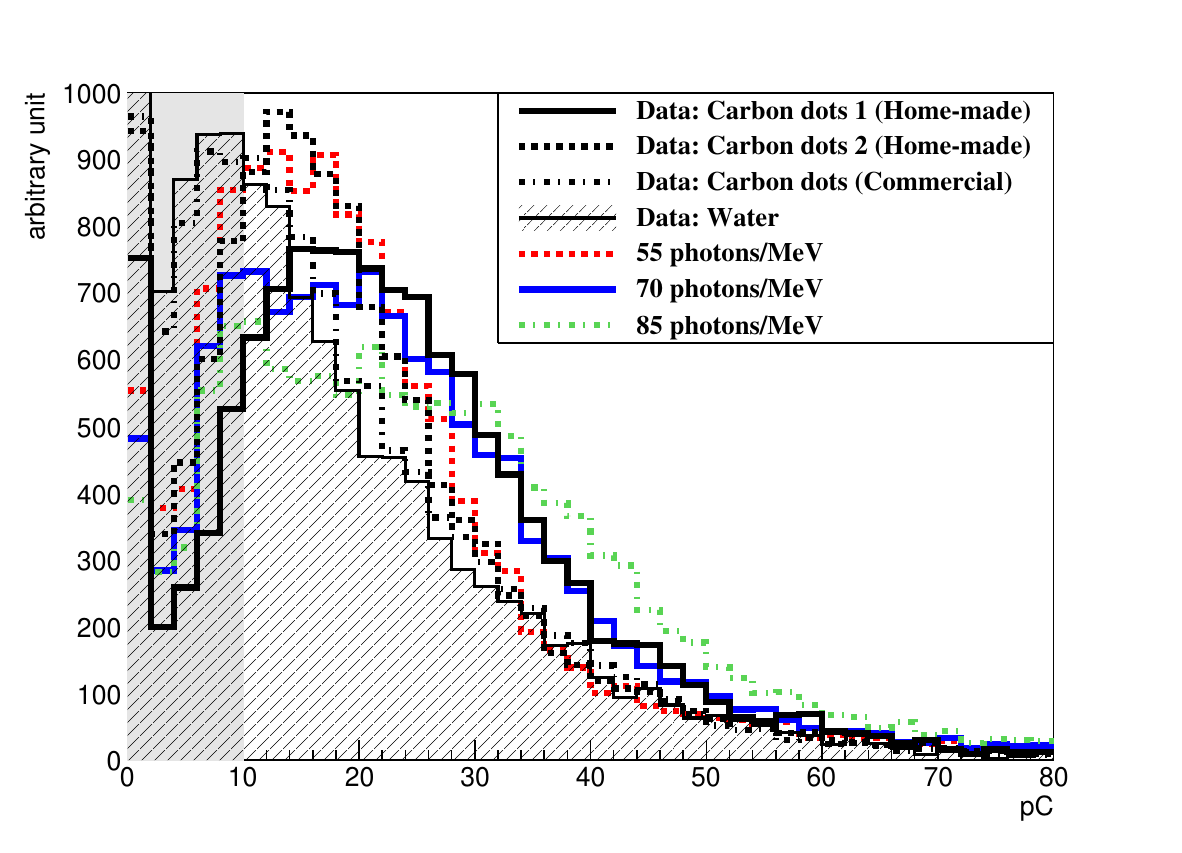}
 \end{center}
\vspace{-2mm}
\caption{Left, a simulation set up. 10 muon trajectories are shown. A hemispherical object models the PMT photo-sensitive region. Right, data-simulation comparison of the charge distribution. There are 4 data samples in black, 2 of carbon dots water dispersion samples (solid and dashed), a commercial carbon dots water dispersion (dotted) and water (grey hatched region). Colored lines are the simulation. Grey shaded region is defined low hits region where the simulation cannot model properly.}
\label{fig:sim}
\end{figure}

To extract photon emission information from these data, we constructed a Geant4-based simulation~\cite{AGOSTINELLI2003250}. Dimensions of the experimental configuration were taken by measurements to make the simulation geometry shown in Fig.~\ref{fig:sim}, left. 
For PMT, only hemi-spherical photo-sensitive region was modelled, and typical quantum efficiency of photo-cathode ~\cite{Hamamatsu} with peak value of $\qe\pm\qee$\% and $\resolpmt$\% resolution was included. 
 4~GeV muons were uniformly distributed on the top taggers with angles that ensure muon passing through the bottom tagger. 
Optical and physical properties of the carbon dots water-based liquid scintillator were modelled as water. Attenuation is not included as the sample is small, and $\resolscat$\% smearing is added to represent scattering effect. Scintillation emission spectrum of the carbon dots was unknown, and we modelled it from our measured photo-luminescence spectra shown in Fig.~\ref{fig:optics}, left. This is smeared $\resolflour$\%, to take into account broadening of the scintillation spectrum. We look to find the scintillation yield, the number of generated photons per deposited energy, by comparing the data with this simulation.

Fig.~\ref{fig:sim}, right, is our main result. Here, four different data sets (black lines) are compared with three different simulations. They are relatively normalized to allow comparison of their shapes.  First, data of two home-made carbon dots samples are shown in black solid and dashed lines, and they are compared with the commercial carbon dots sample (black dotted line). These are qualitatively different shapes from the water-only data (black hatched area), and the data reveal the presence of scintillation light. 
Note, commercial carbon dots water dispersion exhibited the weakest scintillation signal which was due to its low concentration ($<0.01$~\% in distilled water). 
They are compared with three simulation results. 
Many simulated events contain low hits, that are not reflected in the experimental data; to counteract this issue, we cut events less than $\photoncut$ photons from processes other than scintillation (the gray shaded area in Fig.~\ref{fig:sim}). Thus, we focus on a high charge region defined by the total charge to be $>\datapC$ pC for this analysis. In this region, water data agree with the simulation without scintillation light. 

Although the simulation cannot reproduce the overall shape of the data very well, from the data-simulation comparisons, we estimate scintillation yields of home-made carbon dots water-based liquid scintillator can achieve up to $\nphoto\pm\nphotoe$~/MeV, corresponding to the scintillation emission peak between $\photol-\photom$~pC. The error represents the given sample variations and the PMT performance. Although the measurement is not accurate, it shows a reasonable scintillation light from carbon dots water liquid scintillator samples. This is the first demonstration of home-made carbon dots water-based liquid scintillators used in radiation detection.

\section{Outlook
\label{sec:conclusion}
}

In this work, the performance of home-made carbon dots as a water-based liquid scintillator was investigated. The emission behaviour of the home-made samples was comparable to that of the commercial sample, suggesting that the underlying emissive states were similar; however, the intensity of emission was overall lower, indicating a lower emission efficiency of home-made samples. Dynamic light scattering measurements showed multimodal size distributions for the home-made samples, in contrast to the single-peak distribution of the commercial sample. Together with the formation of filament-like residue over time, this indicates that the home-made samples contained additional aggregated or residual byproduct populations, consistent with the absence of post-synthesis purification. These observations indicate that the present optical limitations of samples made at home by the microwave-assisted synthesis could be mitigated through synthesis optimization and addition of purification steps.

The scintillation yield measured with atmospheric muons for the carbon dot dispersions was of order $\nphoto$~MeV$^{-1}$, demonstrating that these samples produce detectable scintillation light in water. While the yield is lower than that of conventional liquid scintillators, it is comparable to mineral-oil-based MiniBooNE neutrino detector~\cite{MiniBooNE:2008paa,rex}. This opens the door to using carbon dots as an alternative, environmentally and biologically safe and inexpensive medium to detect hadronic activity, including low-energy protons in water Cherenkov neutrino detectors. 

As an example, the carbon dots samples produced in this work cost only $\sim$\$0.02$/$L; scaling this to a detector volume comparable to Hyper-Kamiokande (260~kton)~\cite{Hyper-Kamiokande:2018ofw,Hyper-Kamiokande:2025fci} would correspond to a total cost $\sim$\$3.5~M. Such a hypothetical water Cherenkov neutrino detector could enable the tagging low-energy protons to understand nucleon correlations and final state interactions in neutrino interaction physics~\cite{Katori:2016yel,NuSTEC:2017hzk}, the utilization of calorimetric energy reconstruction method for low-energy protons to measure neutral-current quasielastic scattering cross-sections~\cite{MiniBooNE:2010xqw,MiniBooNE:2013dds} and the strange quark contribution on a proton spin~\cite{KamLAND:2022ptk}, the estimation of the tau neutrino fraction in atmospheric neutrinos from the low-energy hadron shower measurements~\cite{Li:2016kra,Super-Kamiokande:2017edb} and the search for new physics with the neutrino flavour triangle~\cite{Wen:2023ijf}, the diffuse supernova neutrino background searches with proton responses instead of neutrons~\cite{Super-Kamiokande:2024kcb,Super-Kamiokande:2025sxh}, the extension of the boosted WIMP dark matter search down to a lower mass region~\cite{MiniBooNEDM:2018cxm,CCM:2021yzc,Super-Kamiokande:2022ncz}, etc. It is also intriguing to consider such environmentally safe and inexpensive technology for neutrino detection in fission nuclear reactor monitoring
~\cite{KamLAND-Zen:2024eml,SNO:2025chx,JUNO:2025gmd,Bernstein:2019hix,Bernstein:2008tj,Classen:2015byu,NUCIFER:2015hdd,DoubleChooz:2019qbj,DayaBay:2022orm,RENO:2020dxd,NEOS:2016wee,PROSPECT:2022wlf,STEREO:2022nzk,NEUTRINO-4:2018huq,Oguri:2014gta,NuLat:2015wgu,Haghighat:2018mve,SoLid:2020cen,Goldsack:2022vzm}. 
Currently, however, the scintillation yield is too low for reactor neutrino detection via inverse beta decay, and further improvements are required.

Several challenges remain for the realistic application of carbon dot water-based liquid scintillators in radiation detection. First, the colloidal properties must be better controlled, in particular through improved synthesis and post-synthesis purification, to suppress aggregation and remove residual byproducts. Aggregation reduces emission efficiency, accelerates ageing, and may pose challenges for filtration and circulation in large-scale water systems. A well-controlled production procedure, including optimisation of concentration, will be required for scalable deployment. Second, the optical properties of the medium, in particular attenuation, need to be characterised, as this will determine the balance between light yield and transparency. Third, long-term stability requires further study; in addition to suppressing aggregation, appropriate surface functionalisation may be necessary to stabilise the carbon dots over extended periods. Finally, material compatibility with ultra-pure water systems must be established. Addressing these challenges through improved synthesis and purification will enable carbon dot dispersions to realise their potential as a scalable water-based scintillation medium.

\section*{Acknowledgment}

We thank members of the Experimental Particle and Astroparticle Physics group at King’s College London for their assistance. We thank Megan Grace-Hughes for producing the image in Fig~\ref{fig:cartoon}. We acknowledge the contributions of Laua Sugii and Subin Chai to the production of carbon dots, and thank Robert Kralik, Daniel Cookman, Gianmaria Collazuol, and Rex Tayloe for valuable discussions and input.

\bibliographystyle{JHEP}
\bibliography{main}

\end{document}